\documentclass[12pt,a4paper]{article}
\usepackage{graphicx}
\usepackage{wrapfig}
\begin{document}
\begin{center}
{\bfseries   Fate of Light Scalar Mesons \footnote{{\small Invited
talk at the International  Conference "Problems of Theoretical and
Mathematical Physics" dedicated to the 110th anniversary of the
birth of the outstanding Russian scientist - mathematician and
physicist Nikolai Nikolaevich Bogolyubov, JINR, Dubna, September
11 - 13, 2019.
 }}} \vskip 5mm N.N. Achasov \vskip 5mm {\small {\it
Sobolev Institute for Mathematics, 630090 Novosibirsk, Russia\\
email: achasov@math.nsc.ru }}
\end{center}

\vskip 4mm \centerline{\bf Abstract} \vskip 1mm It is shown that
all predictions for the light scalars, based on their four-quark
nature, are supported by experiment. The future research program
is outlined also. \\[18pt] PACS numbers: PACS numbers: 12.39.-x,
13.40.-f, 13.60.  Le, 13.75. Lb

 \vskip 8mm

\noindent\textbf{1\ \ Introduction}
\begin{center}
{\bf OVERTURE}
\end{center}
\vspace*{2mm}
 Peter Higgs, C. R. Physique 8 (2007) 970-972:\\[6pt]
 " Another example, which comes closer to the
kind of symmetry breaking which is of interest in particle
physics, is superfluidity. In 1947 Bogoliubov  studied Bose
condensation of an infinite system of neutral spinless bosons with
short-range repulsive two-body interactions. Such a condensate is
characterised by a 'macroscopic wave function' (the order
parameter) which is complex; its modulus squared is a measure of
the observable condensate density, but its argument (which is
unobservable) is arbitrary, thus breaking the symmetry of the
dynamics under rotations of the boson wave functions in the Argand
diagram. The short-range interactions are represented in the
second-quantised Hamiltonian by a term proportional to the square
of the particle density, that is, to a quartic in the components
of the scalar quantum field."\\[6pt]
 See also, N.N. Achasov, Physics of Particles and Nuclei, 2010, Vol. 41, No. 6, pp.
 891-895.\\
 DOI: 10.1134/S1063779610060134, arXiv:1001.3468 [hep-ph].

 \vspace*{6mm}

\noindent\textbf{2 Outline}\\

 \vspace*{3mm}

  The $a_0(980)$ and
$f_0(980)$ mesons are well-established parts of the proposed light
scalar meson nonet \cite{PDG}.  From the beginning, the $a_0(980)$
and $f_0(980)$ mesons became one of the central problems of
nonperturbative QCD, as they are important for understanding the
way chiral symmetry is realized in the low-energy region and,
consequently, for understanding confinement. Many experimental and
theoretical papers have been devoted to this subject.  There is
much evidence that supports the four-quark model of light scalar
mesons \cite {Ja77,SW}.

 The suppression of the $a^0_0(980)$ and
$f_0(980)$ resonances in the $\gamma\gamma\to\eta\pi^0$ and
$\gamma\gamma\to\pi\pi$ reactions, respectively, was predicted in
1982 \cite{ADSh82}, $\Gamma_{a^0_0\gamma\gamma}\approx
\Gamma_{f_0\gamma\gamma}\approx 0.27$ keV, and confirmed by
experiment \cite{PDG}.

 The high quality Belle data \cite{MoUe070809}, Fig. 1, 2,
 allowed to elucidate the mechanisms
of the $\sigma(600)$, $f_0(980)$, and $a^0_0(980)$ resonance
production in $\gamma\gamma$ collisions confirmed their four-quark
structure. $\sigma(600)=f_0(500)$!

 Light scalar mesons are
produced in $\gamma\gamma$ collisions via rescatterings, mainly
via the $\gamma\gamma\to\pi^+\pi^-\to f_0(500)$, $\gamma\gamma\to
K^+K^-\to f_0(980)/a_0(980$ transitions, that is, via the
four-quark transitions. As for $a_2(1320)$ and $f_2(1270)$ (the
well-known $q\bar q$ states), they are produced mainly via the
two-quark transitions (direct couplings with $\gamma\gamma$)
\cite{08,09,11,18}.

 As a
result the practically model-independent prediction of the $q\bar
q$ model $g^2_{f_2\gamma\gamma}:g^2_{a_2\gamma\gamma}=25:9$ agrees
with experiment rather well. As to the ideal  $q\bar q$ model
prediction $g^2_{f_0\gamma\gamma}:g^2_{a_0\gamma\gamma}=25:9$, it
is excluded by experiment.
%\vspace{3mm}
\begin{figure}[h]
\centering \vspace*{-3mm}
\includegraphics[width=36pc,height=17pc]{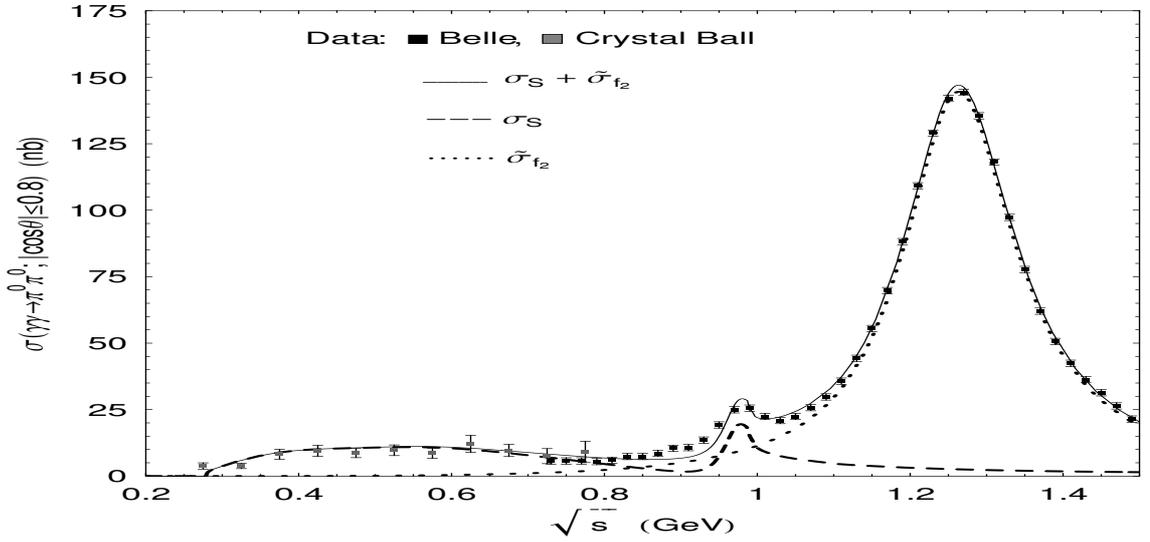}
\caption{ The Belle data \cite{MoUe070809}. Our fit from Ref.
\cite{08}}\label{fig1}
\end{figure}
\vspace*{-2pt}
\begin{figure}[h]
\centering \vspace*{-3mm}
\includegraphics[width=36pc,height=17pc]{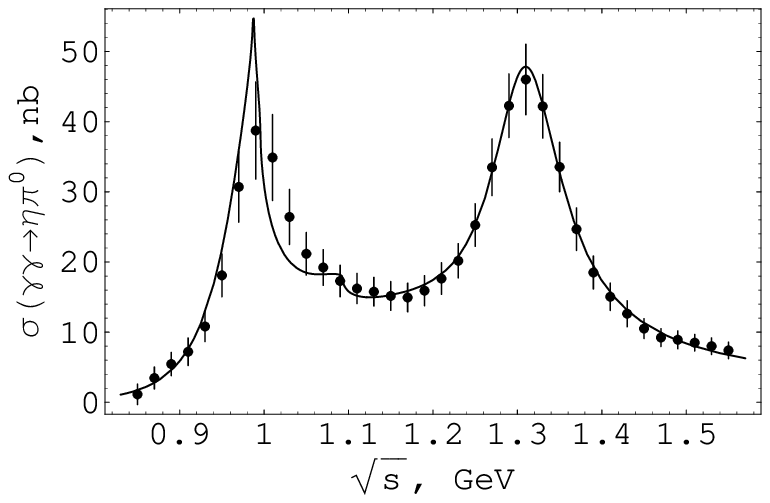}
\caption{The Belle date \cite{MoUe070809}. Our fit from
 Ref. \cite{18}.}\label{fig2}
\end{figure}
\newpage
 Note also that the absence of $J/\psi\to \gamma
f_0(980),\ \rho a_0(980),\ \omega f_0(980)$ decays in the presence
of the intense $J/\psi\to \gamma f_2(1270),\ \gamma f'_2(1525),\\
\rho a_2(1320),\ \omega f_2(1270)$ decays is at variance with the
$P$-wave two-quark, $q\bar q$, structure of $a_0(980)$ and
$f_0(980)$ resonances \cite{98,2002}.

The argument in favor of the four-quark nature of $a_0(980)$ and
$f_0(980)$ is the fact that the $\phi(1020)\to a^0_0(980)\gamma$
and $\phi(1020)\to f_0(980)\gamma$ decays go through the kaon
loop: $\phi\to K^+K^-\to a^0_0(980)\gamma$, $\phi\to K^+K^-\to
f_0(980)\gamma$, i.e., via the four-quark transition
\cite{89,97,01,03,NNA03,06,12,18}.

 The kaon-loop model was suggested in
Ref. \cite{89}  and confirmed by experiment ten years later
\cite{SND,CMD,EOLK,KLOE}, Figs. 3, 4.

 In Ref. \cite{NNA03} it was shown
   that the production of
$a^0_0(980)$ and $f_0(980)$ in $\phi\to a^0_0(980)\gamma\to
\eta\pi^0\gamma$ and $\phi\to f_0(980) \gamma\to\pi^0\pi^0\gamma$
decays is caused by the four-quark transitions, resulting in
strong restrictions on the large $N_C$ expansion  of the decay
amplitudes. The analysis showed that these constraints give new
evidence in favor of the four-quark nature of the $a_0(980)$ and
$f_0(980)$ mesons.

\begin{figure}[h]
\centering \vspace*{-3mm}
\includegraphics[width=36pc,height=15.5pc]{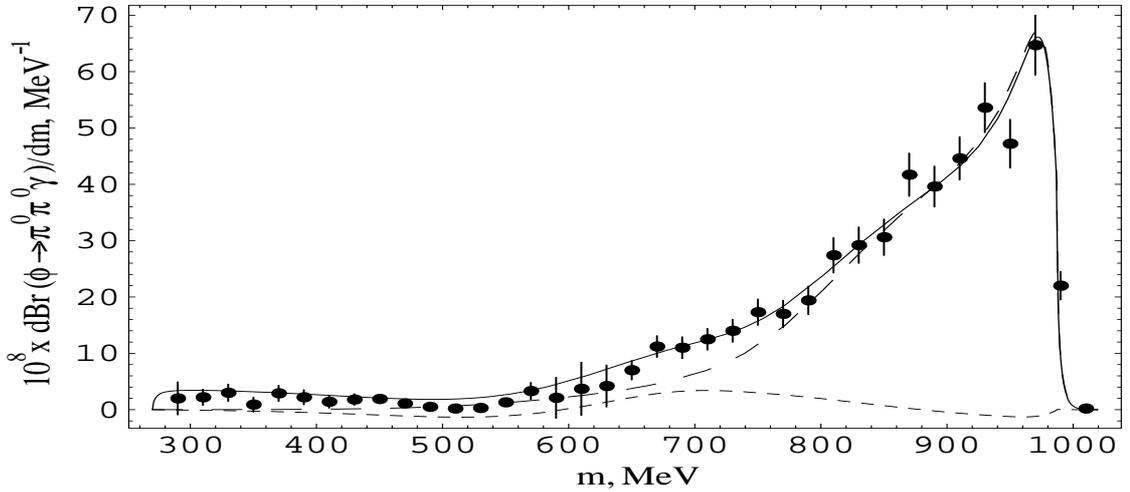}
\caption{The KLOE data \cite{EOLK}. Our fit  Ref.
\cite{12}.}\label{fig3}
\end{figure}

\begin{figure}[h]
\centering \vspace*{-3mm}
\includegraphics[width=36pc,height=15.5pc]{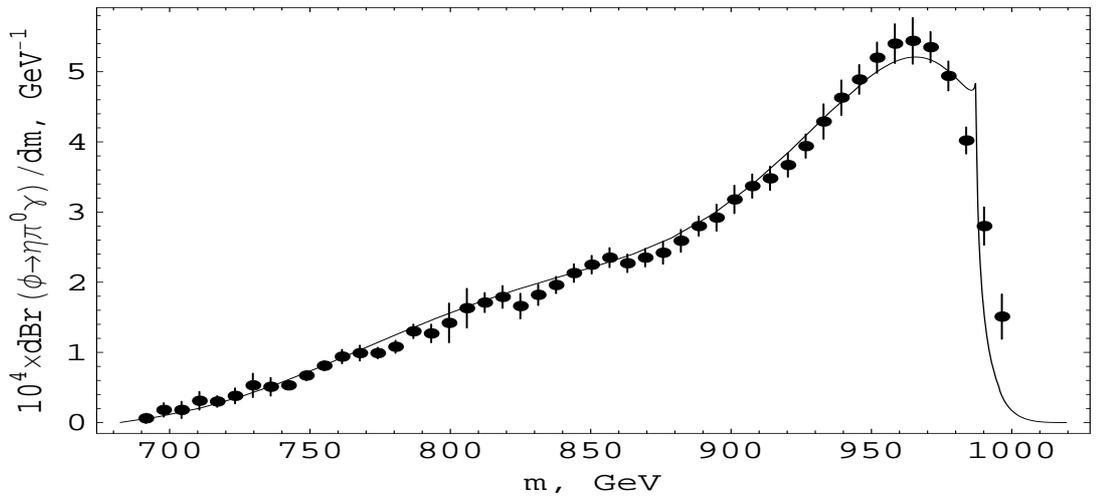}
\caption{The KLOE data \cite{KLOE}. Our fit from
 Ref. \cite{18}.}\label{fig4}
\end{figure}

\newpage

In Refs. \cite{AK0708,AGShev}   it was shown that the description
of the $\phi\to K^+K^-\to\gamma a^0_0(980)/f_0(980)$ decays
requires virtual momenta of $K (\bar K)$ greater than $2$ GeV,
while in the case of loose molecules with a binding energy about
20 MeV, they would have to be about 100 MeV. Besides, it should be
noted that the production of scalar mesons in the pion-nucleon
collisions with large momentum transfers also points to their
compactness \cite{AS98}.

In Refs. \cite{AS9407} it was also shown   that the linear
$S_L(2)\times S_R(2)$ $\sigma$ model \cite{GL60} contains a chiral
shielding of the $\sigma$ meson and reflects all of the main
features of low energy $\pi\pi\to\pi\pi$ and
$\gamma\gamma\to\pi\pi$ reactions up to energy 0.8 GeV and agrees
with the four-quark nature of the $\sigma$ meson.

 This allowed for the development of a
phenomenological model with the right analytical properties in the
complex $s$ plane that took into account the linear $\sigma$
model, the $\sigma(600)-f_0(980)$ mixing and the background
\cite{AK1112}.

\begin{figure}[h]
\centering \vspace*{-3mm}
\includegraphics[width=36pc,height=11.5pc]{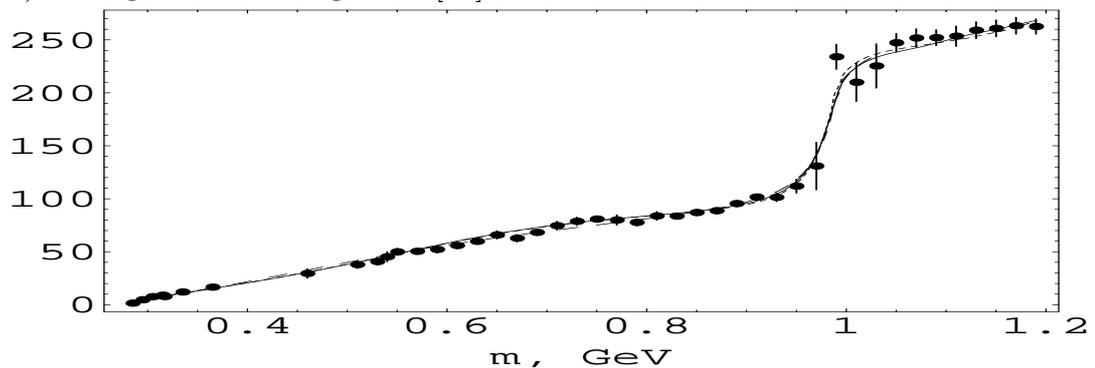}
\caption{The phase $\delta^0_0$ of the $\pi\pi$ scattering
(degrees) is shown.}\label{fig5}
\end{figure}

 This background has a left cut inspired by
crossing symmetry, and the resulting amplitude agrees with results
obtained using the chiral expansion, dispersion relations, and the
Roy equation \cite{CCL} and with the four-quark nature of the
$\sigma(600)$ and $f_0(980)$ mesons as well. This model well
describes the experimental data on $\pi\pi\to\pi\pi$ scattering up
to 1.2 GeV, Fig. 5.

 In Refs. \cite{18,CORREL} It is shown   that
the recent data on the $K^0_S K^+$ correlation in Pb-Pb
interactions Ref. \cite{ALICE}, Fig. 6,
 agree with the data on the
$\gamma\gamma\to\eta\pi^0$ and $\phi\to\eta\pi^0\gamma$ reactions
and support the four-quark model of the $a_0(980)$ meson. It is
shown that the data does not contradict the validity of the
Gaussian assumption.
 %\vspace*{3mm}
 \begin{figure}[h]
\centering \vspace*{-3mm}
\includegraphics[width=36pc,height=11.5pc]{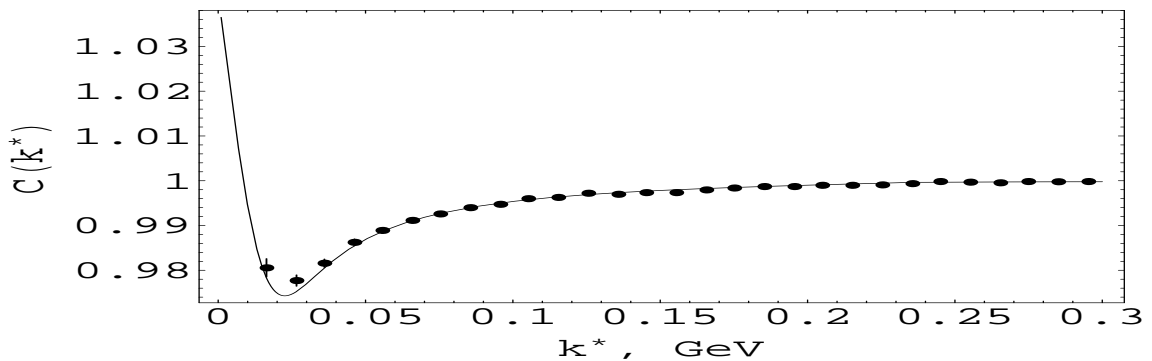}
\caption{The $K_s^0K^+$ correlation $C(k*)$, $k^*$ is the kaon
momentum in the kaon pair rest frame,see Refs.
\cite{18,CORREL}\label{fig6}}
\end{figure}

\newpage
\vspace*{-6mm}
 \noindent\textbf{3 Outlook}\\
  \vspace*{6mm}

 In Refs. \cite{AK12,AK14}it was suggested the program of
studying light scalars in semileptonic $D$ and $B$ decays, which
are the unique probe of the $q\bar q$ constituent pair in the
light scalars. We studied the CLEO data about production of
scalars $\sigma(600)$ and $f_0(980)$ in the $D_s^+\to s\bar s\,
e^+\nu_e\to\pi^+\pi^-\, e^+\nu$ decays, Fig. 7. The conclusion was
that the fraction of the $s\bar s$ constituent components in
$\sigma(600)$ and $f_0(980)$ is small. Unfortunately,  the CLEO
statistics \cite {CLEO} is rather poor, Fig. 8, and thus new
high-statistics data are highly desirable.
 \begin{figure}[h]
\centering \vspace*{-3mm}
\includegraphics[width=36pc,height=15pc]{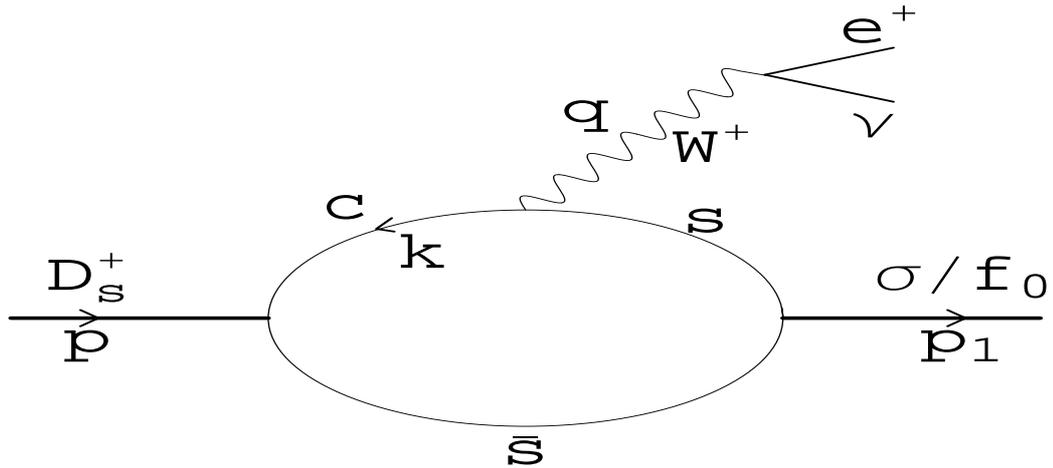}
\caption{The mechanisms of the $D^+_s\to\sigma/f_0\, e^+\nu$
 decays.}\label{fig7}
\end{figure}

 \begin{figure}[h]
\centering \vspace*{-3mm}
\includegraphics[width=36pc,height=15pc]{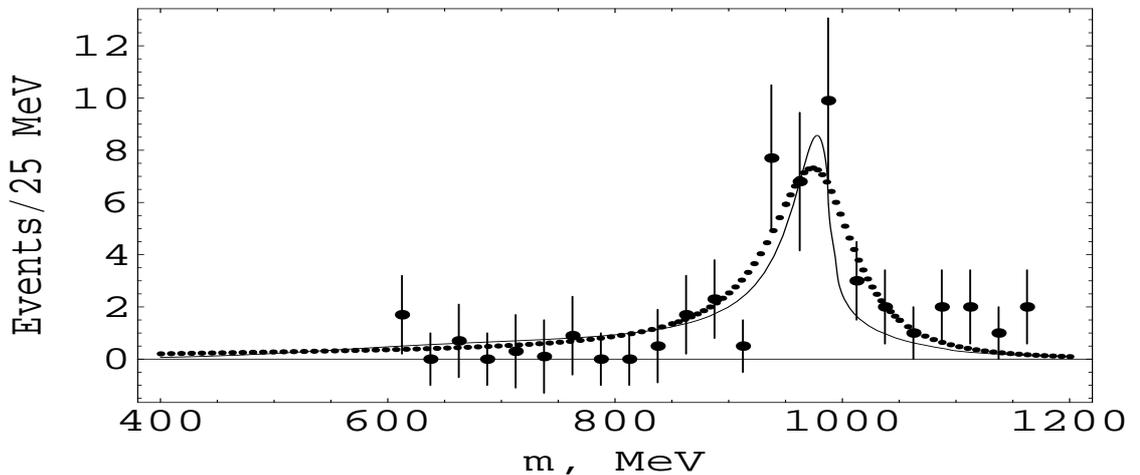}
\caption{The CLEO data  on the invariant $\pi^+\pi^-$ mass ($m$)
distribution for $D^+_s\to\pi^+\pi^-e^+\nu$ decay. The dotted line
is the fit from CLEO. Our theoretical curve is the solid
line.}\label{fig8}
\end{figure}

In Refs. \cite{AK12,AK14} it was noted that no less interesting is
the study of semileptonic decays of $D^0$ and $D^+$ mesons:\\
$D^+\to d\bar d\, e^+\nu_e\to [\sigma(600)+f_0(980)]e^+\nu_e\to
\pi^+\pi^-e^+\nu_e$,\\ $D^0\to d\bar u\, e^+\nu_e\to
a_0^-e^+\nu_e\to\pi^-\eta e^+\nu_e$ and\\ $D^+\to d\bar d\,
e^+\nu_e\to a_0^0 e^+\nu_e\to\pi^0\eta e^+\nu_e$\\ or the
charged-conjugated ones which had not been investigated.

It is  tempting to study light scalar mesons in semileptonic
decays of $B$ mesons \cite{AK14}:\\ $B^0\to d\bar u\, e^+\nu_e\to
a_0^-e^+\nu_e\to\pi^-\eta e^+\nu_e$,\\ $B^+\to u\bar u\,
e^+\nu_e\to a_0^0 e^+\nu_e\to\pi^0\eta e^+\nu_e$ and\\ $B^+\to
u\bar u\, e^+\nu_e\to [\sigma(600)+f_0(980)]e^+\nu_e\to
\pi^+\pi^-e^+\nu_e$\\ or the charged-conjugated ones.

 Recently BESIII Collaboration measured the
decays $D^0\to d\bar u\, e^+\nu\to a_0^-e^+\nu\to\pi^-\eta e^+\nu$
and $D^+\to d\bar d\, e^+\nu\to a_0^0 e^+\nu\to\pi^0\eta e^+\nu$
for the first time \cite{BESIII}.

In Ref. \cite{18}  we discuss these measurements taking into
 account also contribution of $a_0'$ meson with mass about $1400$ MeV, Fig. 9 and Fig. 12 .

\begin{figure}[h]
\centering \vspace*{-3mm}
\includegraphics[width=36pc,height=15pc]{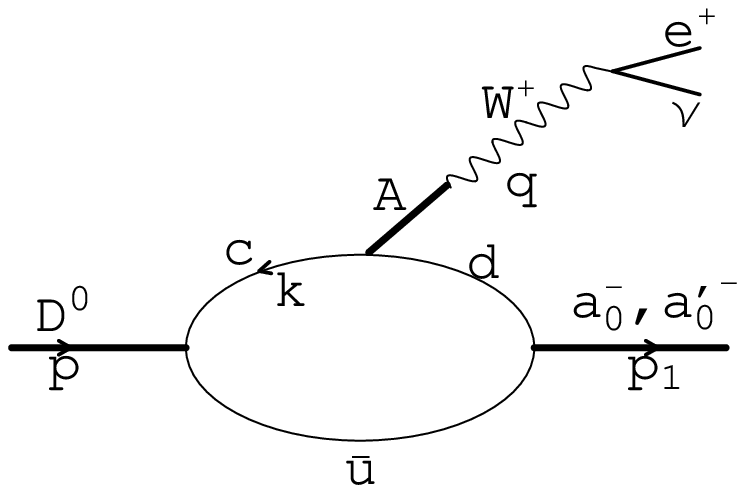}
\caption{The mechanisms of the $D^0\to d\bar u\, e^+\nu_e\to
a_0^-,\, a_0^{\prime -}e^+\nu_e$ decays.}\label{fig9}
\end{figure}

 Below, Figs. 10, 11, is presented a variant when $a_0^-(980)$
has no $q\bar q$ constituent component at all,  that is,
$a_0^-(980)$ is produced as a result of mixing
$a_0^{\prime-}(1400)\to a_0^-(980)$,
 $D^0\to d\bar u\, e^+\nu_e\to a_0^{\prime -}e^+\nu_e \to a_0^-e^+\nu_e\to \pi^-\eta e^+\nu_e$.
  \begin{figure}[h]\centering \vspace*{-3mm}
\includegraphics[width=36pc,height=15pc]{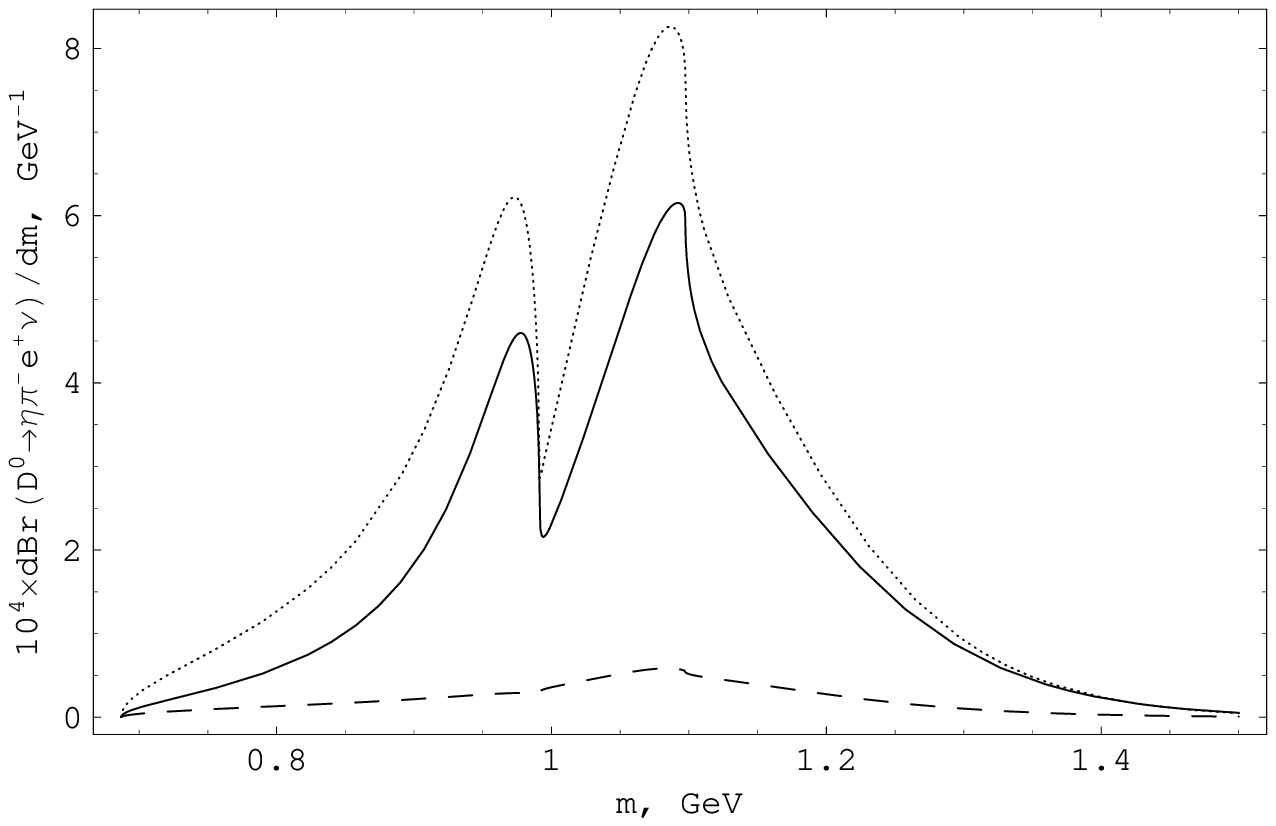}
\caption{The $D^0\to d\bar u\, e^+\nu_e\to a_0^{\prime -}e^+\nu_e
\to a_0^-e^+\nu_e\to \pi^-\eta e^+\nu_e$ spectrum.}\label{fig10}
\end{figure}

 \begin{figure}[h]
\centering \vspace*{-3mm}
\includegraphics[width=36pc,height=12pc]{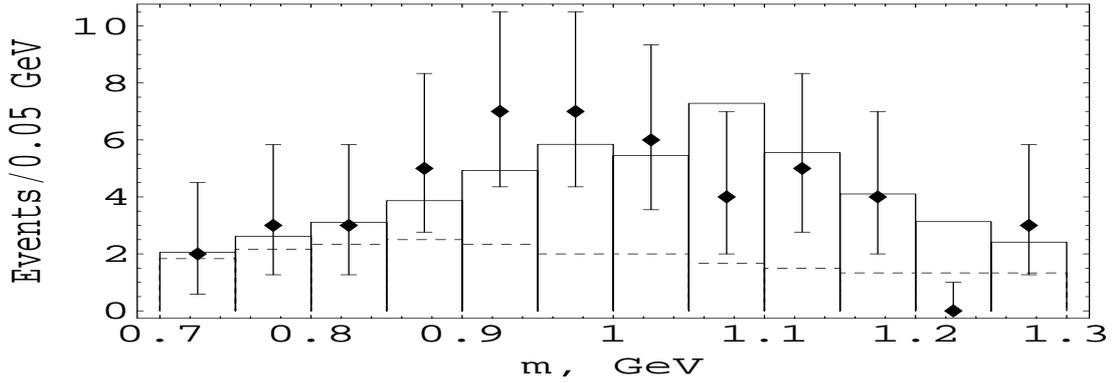}
\caption{The $D^0\to d\bar u\, e^+\nu_e\to a_0^{\prime -}e^+\nu_e
\to a_0^-e^+\nu_e\to \pi^-\eta e^+\nu_e$ spectrum.}\label{fig11}
\end{figure}

\begin{figure}[h]
\centering \vspace*{-3mm}
\includegraphics[width=36pc,height=12pc]{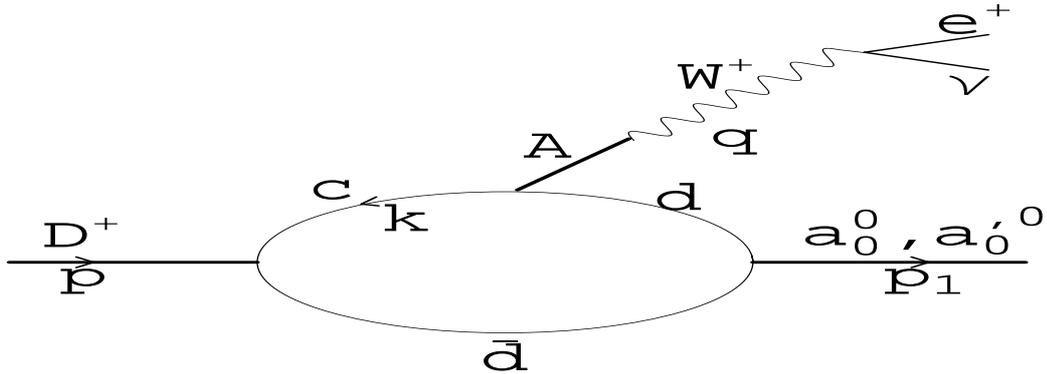}
\caption{The mechanisms of the $D^+\to d\bar d\, e^+\nu_e\to
a_0^0,\, a_0^{\prime 0}e^+\nu_e$ decays.}\label{fig12}
\end{figure}
%\newpage
 Below, Figs. 13, 14, is presented a variant when $a_0^0(980)$
has no $q\bar q$ constituent component at all,  that is,
$a_0^0(980)$ is produced as a result of mixing
$a_0^{\prime0}(1400)\to a_0^0(980)$, $D^+\to d\bar d\, e^+\nu_e\to
a_0^{\prime 0}e^+\nu_e \to a_0^0e^+\nu_e\to \pi^0\eta e^+\nu_e$.
%\vspace*{-12pt}
  \begin{figure}[h] \centering \vspace*{-3mm}
\includegraphics[width=36pc,height=12pc]{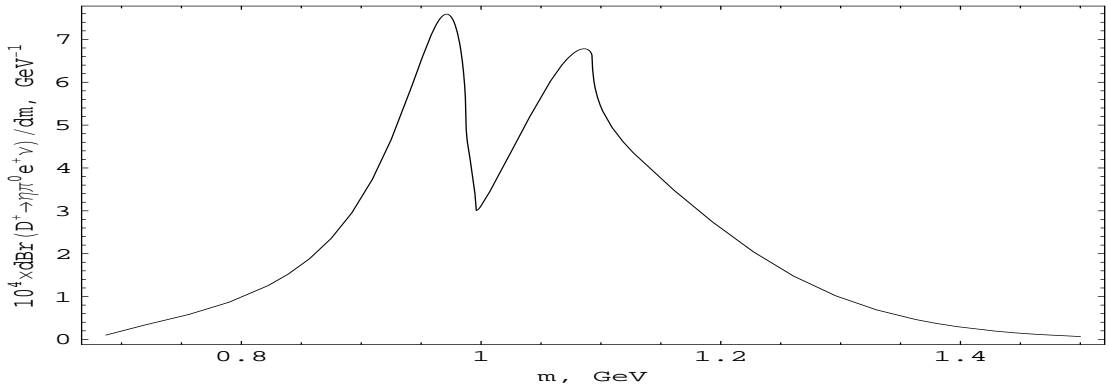}
 \caption{The $D^+\to d\bar d\, e^+\nu_e\to a_0^{\prime
0}e^+\nu_e \to a_0^0e^+\nu_e\to \pi^0\eta e^+\nu_e$
spectrum.}\label{fig13}
\end{figure}
\newpage

%\vspace*{24pt}

\begin{figure}[h]
\centering \vspace*{-3mm}
\includegraphics[width=36pc,height=15pc]{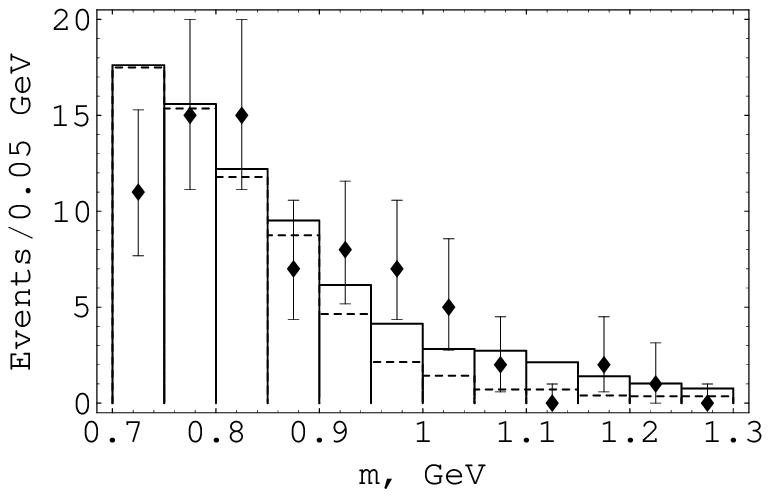}
\caption{The $D^+\to d\bar d\, e^+\nu_e\to a_0^{\prime 0}e^+\nu_e
\to a_0^0e^+\nu_e\to \pi^0\eta e^+\nu_e$ spectrum.}\label{fig14}
\end{figure}
%\newpage
 The first measurements of BESIII directly indicate the absence of
 the constituent $d\bar d$ and $d\bar u$ pairs in the $a_0^0(980)$ and $a_0^-(980)$ states respectively.
  But the the present the statistics is not adequate for the
conclusions about details of the $\pi\eta$ production.

 \vspace*{1cm}
\noindent\textbf{\ \ Acknowledgements }
\vspace{4mm}

 I am grateful
Organizers  for the kind Invitation.

I am grateful to  A.V. Kiselev and G.N. Shestakov for
collaboration.

 The work was supported by the program of fundamental
scientific researches of\\ the SB RAS No. II.15.1., project No.
0314-2019-0021.

%\vspace{8mm}
 \footnotesize{}
\end{document}